\documentclass[12pt]{iopart}
\usepackage{epsfig}

\def\beq{\begin{equation}}
\def\eeq{\end{equation}}

\def\barr{\begin{eqnarray}}
\def\earr{\end{eqnarray}}
\def\lsim{\raise0.3ex\hbox{$\;<$\kern-0.75em%
\raise-1.1ex\hbox{$\sim\;$}}}
\def\gsim{\raise0.3ex\hbox{$\;>$\kern-0.75em%
\raise-1.1ex\hbox{$\sim\;$}}}
\def\bmat{\left( \begin{array}}
\def\emat{\end{array} \right)}

\def\dmsq{\Delta m^2}
\def\msqa{\Delta m^2_{\rm atm}}
\def\msqs{\Delta m^2_{\odot}}

\def\nuebar{{\bar{\nu}_e}}
\def\nux{{\nu_x}}
\def\ebar{{\bar{e}}}

\begin{document}

\title[Supernova Neutrino Oscillations at IceCube]
{Detecting the Neutrino Mass Hierarchy with a Supernova
at IceCube}

\author{Amol S.~Dighe, Mathias Th.~Keil and Georg G.~Raffelt}

\address{Max-Planck-Institut f\"ur Physik
(Werner-Heisenberg-Institut)\\
F\"ohringer Ring 6, 80805 M\"unchen, Germany}

\begin{abstract}
IceCube, a future km$^3$ antarctic ice Cherenkov neutrino telescope,
is highly sensitive to a galactic supernova (SN) neutrino burst.  The
Cherenkov light corresponding to the total energy deposited by the SN
neutrinos in the ice can be measured relative to background
fluctuations with a statistical precision much better than 1\%. If the
SN is viewed through the Earth, the matter effect on neutrino
oscillations can change the signal by more than 5\%, depending on the
flavor-dependent source spectra and the neutrino mixing parameters.
Therefore, IceCube together with another high-statistics experiment
like Hyper-Kamiokande can detect the Earth effect, an observation that
would identify specific neutrino mixing scenarios that are difficult
to pin down with long-baseline experiments.  In particular, the normal
mass hierarchy can be clearly detected if the third mixing angle is
not too small, $\sin^2\theta_{13}\gsim 10^{-3}$.  The small
flavor-dependent differences of the SN neutrino fluxes and spectra
that are found in state-of-the-art simulations suffice for this
purpose.  Although the absolute calibration uncertainty at IceCube may
exceed 5\%, the Earth effect would typically vary by a large amount
over the duration of the SN signal, obviating the need for a precise
calibration.  Therefore, IceCube with its unique geographic location
and expected longevity can play a decisive role as a ``co-detector''
to measure SN neutrino oscillations. It is also a powerful stand-alone
SN detector that can verify the delayed-explosion scenario.
\end{abstract}


\section{Introduction}

The antarctic neutrino telescope AMANDA~\cite{Wiebusch:ij,Amanda} and
the future km$^3$ IceCube~\cite{Ahrens:2002dv,IceCube} are designed to
observe high-energy neutrinos from astrophysical sources. The ice is
instrumented with photomultipliers to pick up the Cherenkov light from
secondary charged particles. In order to reach the large volume needed
to detect the expected small fluxes at high energies, the density of
optical modules is far too sparse to measure, for example, solar
neutrinos.  However, it has been recognized for a long time that these
instruments can detect a supernova (SN) neutrino burst because the
Cherenkov glow of the ice can be identified as time-correlated noise
among all phototubes~\cite{Halzen:xe,Halzen:1995ex}. This approach has
been used by AMANDA to exclude the occurrence of a galactic SN over a
recent observation period~\cite{Ahrens:2001tz}.

For AMANDA the physics potential of a possible SN observation is
essentially limited to its detection, notably in the context of the
Supernova Early Warning System (SNEWS) that would alert the
astronomical community several hours before the optical
explosion~\cite{Scholberg:1999tm,SNEWS}. For the future IceCube with
4800 optical modules, however, the number of detected Cherenkov
photons would be of order $10^6$ and thus so large that several
interesting physics questions could be addressed in earnest.

The observed quantity is the number of Cherenkov photons caused by the
SN neutrinos as a function of time, i.e.\ a measure of the energy
deposited by the neutrinos in the ice. Therefore, the information
about the SN signal is far more limited than what can be extracted
from a high-statistics observation in Super-Kamiokande or other
low-energy experiments that detect individual events.  However,
galactic SNe are so rare, perhaps a few per century, that the chances
of observing one depend crucially on the long-term stability of the
neutrino observatories. Once IceCube has been built it may well
operate for several decades, backing up the low-energy experiments.
Besides the detection and associated early warning one could measure
important details of the neutrino light-curve, for example the
existence and duration of the initial SN accretion phase, the overall
duration of the cooling phase, and so forth. Such an observation would
provide a plethora of astrophysically valuable information.

However, from the perspective of neutrino physics a {\em
simultaneous\/} observation in both IceCube and another large detector
such as Super-Kamiokande or Hyper-Kamiokande would be
especially useful.  Assuming that the neutrinos have traversed
significantly different paths through the Earth, the two signals could
well show measurable differences caused by neutrino oscillations in
matter~\cite{Dighe:1999bi,Dighe:2001rs,Lunardini:2001pb,%
Lunardini:2003eh,Takahashi:2001dc}.  As this Earth effect shows up
only for certain combinations of neutrino mixing parameters, a dual
observation may well distinguish, for example, between the normal and
inverted neutrino mass hierarchy.  It is well known that observing SN
neutrinos with two or more detectors with different Earth-crossing
lengths is extremely useful, but IceCube's potential has not been
explored in this context. With all the low-energy observatories being
in the northern hemisphere, IceCube's location in Antarctica is
uniquely complementary for this purpose.

Any oscillation signature depends on the small flavor-dependent
differences between the fluxes and spectra at the source.  If these
differences were as large as had been assumed until recently there
would be little question about IceCube's usefulness for co-detecting
the Earth effect.  However, a more systematic study of the
flavor-dependence of the SN neutrino fluxes and spectra reveals that
these differences are more subtle, although by no means
negligible~\cite{Raffelt:ai,Buras:2002wt,Keil:2002in,Noon03}.  We
evaluate IceCube's potential as a co-detector from the perspective of
these ``pessimistic'' assumptions about the primary fluxes and
spectra.

This paper is organized as follows. In Sec.~\ref{at-icecube}, we show
that the neutrino signal from a galactic supernova can be measured at
IceCube with a sub-percent statistical precision.  In
Sec.~\ref{earth-effects}, we calculate the Earth matter effects on
this signal and illustrate that it is possible to detect them in
conjunction with another high statistics experiment.  Sec.~\ref{concl}
concludes.

\section{Supernova Neutrino Detection in Ice Cherenkov Detectors}
\label{at-icecube}

\subsection{Cherenkov Photons in One Optical Module}

The SN neutrinos streaming through the antarctic ice interact
according to $\bar\nu_e p\to n e^+$ and some other less important
reactions. The positrons, in turn, emit Cherenkov light that is picked
up by the optical modules (OMs) frozen into the ice.  While the
expected number of detected photons per OM was calculated in
Refs.~\cite{Halzen:xe,Halzen:1995ex}, we revisit their estimate for
two reasons. First, the SN signal was directly scaled to the
historical SN~1987A observation in Kamiokande~II so that the exact
assumptions about the neutrino flux are not directly apparent.
Second, the expected number of Cherenkov photons detected by one OM
was based on estimating an effective ice volume seen by
one~OM. However, it is much simpler to work in the opposite direction
and start with the homogeneous and isotropic Cherenkov glow of the ice
caused by the SN neutrinos.  The OM is immersed in this diffuse bath
of photons and picks up a number corresponding to its angular
acceptance and quantum efficiency.

As a first simplification we limit ourselves to the signal caused by
the inverse $\beta$ reaction $\bar\nu_e p\to n e^+$. The $\bar\nu_e$
fluence (time-integrated flux) at Earth is
\begin{equation}
{\cal F}_{\bar\nu_e}=1.745\times10^{11}~{\rm cm}^{-2}\,f_{\rm SN}\,.
\end{equation}
We define the ``SN fudge factor'' as
\begin{equation}
f_{\rm SN}\equiv
\frac{E_{\bar\nu_e,{\rm tot}}}{5\times10^{52}~{\rm erg}}
~\frac{15~{\rm MeV}}{\langle E_{\bar\nu_e}\rangle}
~\left(\frac{10~{\rm kpc}}{D}\right)^2\,,
\end{equation}
where $E_{\bar\nu_e,{\rm tot}}$ is the total energy leaving the SN in
the form of $\bar\nu_e$ after flavor oscillations have been included,
$\langle E_{\bar\nu_e}\rangle$ is the average $\bar\nu_e$ energy, and
$D$ the distance.

The energy deposited in the ice per target proton is ${\cal
F}_{\bar\nu_e}\,\langle E_{\bar\nu_e}\sigma\rangle$.  For the inverse
$\beta$ cross section we ignore weak-magnetism and recoil corrections
and also the difference between $\bar\nu_e$ and positron energy so
that~\cite{Vogel:1999zy}
\begin{equation}
\sigma=9.52\times10^{-44}~{\rm cm}^2~
\left(\frac{E_{\bar\nu_e}}{\rm MeV}\right)^2\,.
\end{equation}
For the neutrino flux of each neutrino and anti-neutrino species we
assume a distribution of the form~\cite{Keil:2002in}
\begin{equation}\label{eq:spectralform}
F(E)=
\frac{\Phi_0}{E_0}\,\frac{(1+\alpha)^{1+\alpha}}{\Gamma(1+\alpha)}  
\left(\frac{E}{E_0}\right)^\alpha 
\exp\left[-(\alpha+1)\frac{E}{E_0}\right]\,,
\end{equation}
where $E_0$ is the average energy, $\alpha$ a parameter that typically
takes on values 2.5--5 depending on the flavor and the phase of
neutrino emission, and $\Phi_0$ the overall flux at the detector in
units of $\rm cm^{-2}~s^{-1}$. This distribution implies
\begin{equation}
\langle E_{\bar\nu_e}^3\rangle
=\frac{(3+\alpha)(2+\alpha)}{(1+\alpha)^2}\,
\langle E_{\bar\nu_e}\rangle^3
=\frac{15}{8}\,\langle E_{\bar\nu_e}\rangle^3
\hbox{~for $\alpha=3$}\,.
\end{equation}
Altogether we thus find
\begin{equation}
\langle E_{\bar\nu_e}\sigma\rangle=
6.024\times10^{-40}~{\rm MeV}~{\rm cm}^2
~f_\sigma
\end{equation}
with
\begin{equation}
f_\sigma\equiv 
\frac{8\,(3+\alpha)(2+\alpha)}{15\,(1+\alpha)^2}~
\left(\frac{\langle E_{\bar\nu_e}\rangle}{\rm 15~MeV}\right)^3\,.
\end{equation}
This fudge factor can also be taken to include deviations from the
simplified energy dependence of the cross section and deviations from
the assumed spectral shape.

The Cherenkov angle for photon emission by a charged particle is
$\cos\Theta=(n\beta)^{-1}$ where $n$ is the medium's refractive index
and $\beta$ the particle's velocity. With $n=1.31$ for ice, neglecting
the $\lambda$-dependence, and $\beta=1$ we have $\Theta=40.2^\circ$.
A particle with unit charge produces Cherenkov photons per unit path
length and per unit wavelength band according to
\begin{equation}
\frac{d^2N_\gamma}{dx\,d\lambda}=
\frac{2\pi\alpha\sin^2\Theta}{\lambda^2}\,,
\end{equation}
where $\alpha=1/137$ is the fine-structure constant.  Assuming that
$n$ and thus $\Theta$ are independent of wavelength we integrate over
$\lambda$ and find
\begin{equation}
\frac{dN_\gamma}{dx}\bigg|_\lambda^\infty
=638~{\rm cm}^{-1}~\frac{300~{\rm nm}}{\lambda}\,.
\end{equation}
Taking the useful wavelength range to be 300--600~nm this translates
into 319 photons per cm pathlength. Taking the positron mean free path
to be 12~cm for an energy of 20~MeV, and taking it to be proportional
to its energy, the number of useful Cherenkov photons per deposited
neutrino energy is
\begin{equation}
\frac{N_\gamma}{E_{\bar\nu_e}}=191~{\rm MeV}^{-1}~f_{\rm Ch}
\end{equation}
with yet another fudge factor $f_{\rm Ch}$.

The density of ice is $0.924~\rm g~cm^{-3}$, corresponding to about
$6.18\times10^{22}~\rm cm^{-3}$ proton targets. Therefore, the SN
neutrinos produce $1.241\times10^{-3}~{\rm cm}^{-3}\,f_{\rm
SN}\,f_{\sigma}\,f_{\rm Ch}$ useful Cherenkov photons per unit volume
of ice. Multiplying this number with the speed of light and dividing
by $4\pi$ gives us the resulting diffuse photon flux in units of $\rm
cm^{-2}~s^{-1}~ster^{-1}$. However, the average lifetime of these
photons is $c R_{\rm abs}$ with $R_{\rm abs}$ the absorption length.
Therefore, the neutrino-induced photon fluence is found by multiplying
the flux with $c R_{\rm abs}$,
\begin{equation}
\frac{d{\cal F}_\gamma}{d\Omega}=0.9874~{\rm cm}^{-2}~{\rm ster}^{-1}
~f_{\rm SN}\,f_{\sigma}\,f_{\rm Ch}\,f_{\rm abs}
\end{equation}
where $f_{\rm abs}=R_{\rm abs}/100\,{\rm m}$.

The number of events produced by this fluence in a given OM depends on
the average quantum efficiency taken to be $Q=0.20$. In addition, it
depends on the angular acceptance, i.e.\ the effective photo
cathode detection area $A_{\rm cat}$ times the angular acceptance
range $\Omega_{\rm acc}$. Therefore, in one OM we expect
\begin{equation}\label{eq:photoncounts}
N_{\rm events}=310
~f_{\rm SN}\,f_{\sigma}\,f_{\rm Ch}\,f_{\rm abs}\,f_{\rm OM}
\end{equation}
with
\begin{equation}
f_{\rm OM}=\frac{Q}{0.20}
~\frac{A_{\rm cat}}{250~\rm cm^2}
~\frac{\Omega_{\rm acc}}{2\pi}\,.
\end{equation}
This result is independent of the presence of bubbles in the ice that
scatter the photons. The Cherenkov glow of the ice represents an
isotropic and homogeneous distribution that is not changed by elastic
scattering.

\subsection{Comparing With Previous Work}

In order to compare our result with the one derived in
Ref.~\cite{Halzen:1995ex} we need to translate their assumptions into
our fudge factors. The $\bar\nu_e$ distribution was taken to
follow a Fermi-Dirac spectrum with $T=4$~MeV, implying $\langle
E_{\bar\nu_e}\rangle=12.61$~MeV.  The distance of the SN was taken to
be 10~kpc, and the total energy release was scaled to the
Kamiokande~II signal for SN~1987A. With our choice of the $\beta$
cross section these assumptions correspond to $E_{\bar\nu_e,{\rm
tot}}=3.17\times10^{52}$~erg, i.e.\ to $f_{\rm SN}=0.754$.  These
authors also used a quadratic energy dependence of the cross
section. Integrating over their Fermi-Dirac spectrum they effectively
used $f_\sigma=0.663$. Further, they assumed 3000 useful Cherenkov
photons for 20~MeV deposited energy, i.e.\ effectively $f_{\rm
Ch}=0.785$.  For the absorption length they used 300~m, i.e.\ $f_{\rm
abs}=3$.  Finally, they assumed a quantum efficiency of 25\%, a
cathode area of 280~cm$^2$, and an acceptance range of $2\pi$, i.e.\
$f_{\rm OM}=1.12$. Altogether, we find for these assumptions $N_{\rm
events}=409$ per OM. This compares with 273 in
Ref.~\cite{Halzen:1995ex}, i.e.\ our result is larger by a factor~1.5. 

The result in Ref.~\cite{Halzen:1995ex} was backed up by a detailed
Monte Carlo treatment of the production and propagation of Cherenkov
photons in the AMANDA detector. Therefore, the difference may well
relate to details of the OM acceptance and wavelength-dependent
quantum efficiency and photon propagation. Many of these details will
be different in IceCube where 10~inch photomultiplier tubes and
different regions of ice will be used.  Detailed values for the
detector-dependent fudge factors must be determined specifically for
IceCube once it has been built.  The main difference between the
assumptions in Ref.~\cite{Halzen:1995ex} and our estimate is the
absorption length.  When using AMANDA as a SN observatory a realistic
value was taken to be around 100~m~\cite{Ahrens:2001tz}.  The vast
difference between these estimates is that the former was based on the
measured absorption length in a dust-free region of the ice. For our
further estimates we stick to 100~m as a conservative assumption.

\subsection{Event Rate vs.\ Neutrino Luminosity}

In our derivation we have used the time-integrated neutrino flux,
amounting to the assumption of a stationary situation.  The absorption
time for photons is very small, $\tau_{\rm abs}=R_{\rm
abs}/c=0.33~\mu{\rm s}~R_{\rm abs}/100~{\rm m}$. The SN signal will
vary on time scales exceeding 10~ms. Therefore, the Cherenkov glow of
the ice follows the time-variation of the SN signal without
discernible inertia.  Hence one may replace the neutrino fluence with
a time-dependent flux and $N_{\rm events}$ with an event rate
$\Gamma_{\rm events}$.

Moreover, for our further discussion it will be useful to consolidate
our fudge-factors into one describing the detector response, and
others characterizing the neutrino flux. Therefore, we summarize our
prediction for the event rate per OM in the form
\begin{equation}\label{eq:rateprediction}
\Gamma_{\rm events}=62~{\rm s}^{-1}\,
\frac{L_{\bar\nu_e}}{10^{52}~\rm erg~s^{-1}}
~\left(\frac{10~\rm kpc}{D}\right)^2\,
f_{\rm flux}\,f_{\rm det}
\end{equation}
where $L_{\bar\nu_e}$ is the $\bar\nu_e$ luminosity after flavor
oscillations and
\begin{eqnarray}
f_{\rm flux}&=&
\frac{15~\rm MeV}{\langle E_{\bar\nu_e}\rangle}
~\frac{8\,\langle E_{\bar\nu_e}^3\rangle}{15\,(15~\rm MeV)^3}\,,
\label{Lnuebar}\\
f_{\rm det}&=&f_{\rm Ch} \,\frac{R_{\rm abs}}{100~\rm m} ~\frac{Q}{0.20}
~\frac{A_{\rm cat}}{250~\rm cm^2} ~\frac{\Omega_{\rm acc}}{2\pi}\,.
\end{eqnarray}
Here, $f_{\rm det}$ also includes corrections for the
energy dependence of the $\beta$ cross section.

We stress that our simple estimate of the counting rate primarily
serves the purpose of determining its magnitude relative to the
background. The important feature is that the signal relative to the
background can be determined with a good statistical precision. Of
course, for an absolute detector calibration a detailed modeling would
be necessary. For our present purpose, however, even an uncertainty of
several 10\% in our estimated counting rate is irrelevant.

\subsection{Supernova Signal in IceCube}

IceCube will have 4800~OMs so that one expects a total event
number of $1.50\times10^6$, taking all fudge factors to be unity.
Assuming a background counting rate of 300~Hz per OM over as much as
10~s this compares with a background rate of $1.44\times10^7$.
Assuming Poisson fluctuations, the uncertainty of this number is
$3.8\times10^3$, i.e.\ 0.25\% of the SN signal.  Therefore, one can
determine the SN signal with a statistical sub-percent precision,
ignoring for now problems of absolute detector calibration.

\begin{figure}[ht]
\begin{indented}
\item[]
\epsfig{file=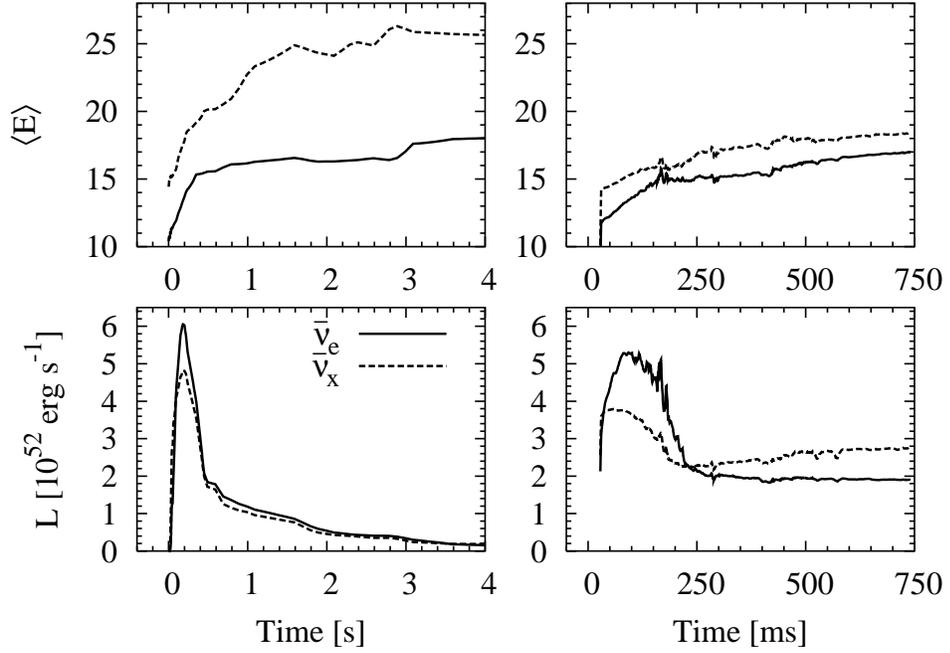,height=9cm}
\end{indented}
\caption{\label{fig:simulations}Supernova $\bar\nu_e$ and $\bar\nu_x$
light curves and average energies.  {\em Left:} Livermore
simulation~\cite{Totani:1997vj}.  {\em Right:} Garching
simulation~\cite{Noon03}.}
\end{figure}

In order to illustrate the statistical power of IceCube to observe a
SN signal we use two different numerical SN simulations.  The first
was performed by the Livermore group~\cite{Totani:1997vj} that
involves traditional input physics for mu- and tau-neutrino
interactions and a flux-limited diffusion scheme for treating neutrino
transport. The great advantage of this simulation is that it covers
the full evolution from infall over the explosion to the
Kelvin-Helmholtz cooling phase of the newly formed neutron star. We
show the Livermore $\bar\nu_e$ and $\bar\nu_x$ lightcurves in
Fig~\ref{fig:simulations} (left panels). Here and in the following we
take $\bar\nu_x$ to stand for either $\bar\nu_\mu$ or
$\bar\nu_\tau$. Apart from very small differences the SN fluxes and
spectra are thought to be equal for $\nu_\mu$, $\nu_\tau$,
$\bar\nu_\mu$, and $\bar\nu_\tau$.

Our second simulation was performed with the Garching
code~\cite{Rampp:2002}.  It includes all relevant neutrino interaction
rates, including nucleon bremsstrahlung, neutrino pair processes, weak
magnetism, nucleon recoils, and nuclear correlation effects.  The
neutrino transport part is based on a Boltzmann solver.  The
neutrino-radiation hydrodynamics program allows one to perform
spherically symmetric as well as multi-dimensional simulations.  The
progenitor model is a 15$\,M_{\odot}$ star with a $1.28\,M_{\odot}$
iron core.  The period from shock formation to 468~ms after bounce was
evolved in two dimensions. The subsequent evolution of the model is
simulated in spherical symmetry.  At 150~ms the explosion sets in,
although a small modification of the Boltzmann transport was necessary
to allow this to happen~\cite{Janka:2002}. Unmanipulated full-scale
models with an accurate treatment of the microphysics currently do not
obtain explosions~\cite{Buras:2003}. This run will be continued beyond
the current epoch of 750~ms post bounce; we here use the preliminary
results currently available~\cite{Noon03}.  We show the Garching
$\bar\nu_e$ and $\bar\nu_x$ lightcurves in Fig~\ref{fig:simulations}
(right panels).

We take the Livermore simulation to represent traditional predictions
for flavor-dependent SN neutrino fluxes and spectra that were used in
many previous discussions of SN neutrino oscillations.  The Garching
simulation is taken to represent a situation when the $\bar\nu_x$
interactions are more systematically included so that the
flavor-dependent spectra and fluxes are more similar than had been
assumed previously~\cite{Raffelt:ai,Buras:2002wt,Keil:2002in,Noon03}.
We think it is useful to juxtapose the IceCube response for both
cases.

Another difference is that in Livermore the accretion phase lasts
longer. Since the explosion mechanism is not finally settled, it is
not obvious which case is more realistic.  Moreover, there could be
differences between different SNe. The overall features are certainly
comparable between the two simulations.

\begin{figure}[ht]
\begin{indented}
\item[]
\epsfig{file=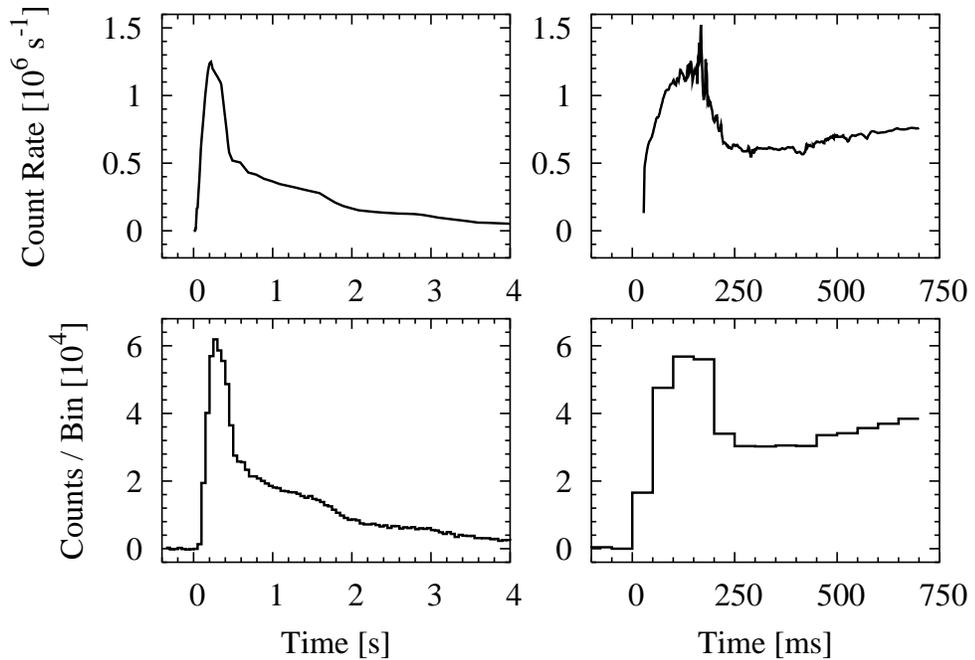,height=9cm}
\end{indented}
\caption{\label{fig:SimulationsIceCube}Supernova signal in IceCube
assuming a distance of 10~kpc, based on the Livermore simulation
(left) and the Garching one (right), in both cases ignoring flavor
oscillations.  In the bottom panels we have used 50~ms bins and have
added noise from a background rate of 300~Hz per OM.}
\end{figure}

In Fig.~\ref{fig:SimulationsIceCube} we show the expected counting
rates in IceCube on the basis of Eq.~(\ref{eq:rateprediction}) for an
assumed distance of 10~kpc and 4800 OMs for the Livermore (left)
and Garching (right) simulations.  We also show this signal in 50~ms
bins where we have added noise from a background of 300~Hz per OM. The
baseline is at the average background rate so that negative counts
correspond to downward background fluctuations.

One could easily identify the existence and duration of the accretion
phase and thus test the standard delayed-explosion scenario. One could
also measure the overall duration of the cooling phase and thus
exclude the presence of significant exotic energy losses.  Therefore,
many of the particle-physics limits based on the SN~1987A
neutrinos~\cite{Raffelt:1999tx} could be supported with a
statistically serious signal.  If the SN core were to collapse to a
black hole after some time, the sudden turn-off of the neutrino flux
could be identified. In short, when a galactic SN occurs, IceCube is a
powerful stand-alone neutrino detector, providing us with a plethora
of information that is of fundamental astrophysical and
particle-physics interest.

In addition, IceCube is extremely useful as a co-detector with another
high-statistics observatory to measure neutrino oscillation effects, a
topic that we now explore.

\newpage

\section{Observing Supernova Neutrinos Through The Earth}
\label{earth-effects}

\subsection{Earth Matter Effect on SN Neutrino Spectra}

Neutrino oscillations are now firmly established by measurements of
solar and atmo\-spheric neutrinos and the KamLAND and K2K long-baseline
experiments~\cite{Gonzalez-Garcia:2002dz,Fogli:2002au%
,Bahcall:2002ij,deHolanda:2002iv,Maltoni:2002aw,Maltoni:2002ni}.
Evidently the weak interaction
eigenstates $\nu_e$, $\nu_\mu$ and~$\nu_\tau$ are non-trivial
superpositions of three mass eigenstates $\nu_1$, $\nu_2$ and~$\nu_3$,
\begin{equation}
\pmatrix{\nu_e\cr \nu_\mu \cr \nu_\tau\cr}
=U\pmatrix{\nu_1\cr \nu_2 \cr \nu_3\cr}\,,
\end{equation}
where $U$ is the leptonic mixing matrix that can be written in the 
canonical form
\begin{equation}
U=\pmatrix{1&0&0\cr0&c_{23}&s_{23}\cr0&-s_{23}&c_{23}\cr}
\pmatrix{c_{13}&0&e^{i\delta}s_{13}\cr0&1&0\cr
-e^{-i\delta}s_{13}&0&c_{13}\cr}
\pmatrix{c_{12}&s_{12}&0\cr-s_{12}&c_{12}&0\cr0&0&1\cr}\,.
\end{equation}
Here $c_{12}=\cos\theta_{12}$ and $s_{12}=\sin\theta_{12}$ etc., and
$\delta$ is a phase that can lead to CP-violating effects, that are,
however, irrelevant for SN neutrinos.

The mass squared differences relevant for the atmospheric and solar
neutrino oscillations obey a hierarchy $\msqa \gg \msqs$. This
hierarchy, combined with the observed smallness of the angle
$\theta_{13}$ at CHOOZ \cite{chooz} implies that the atmospheric
neutrino oscillations essentially decouple from the solar ones and
each of these is dominated by only one of the mixing angles.  The
atmospheric neutrino oscillations are controlled by $\theta_{23}$ that
may well be maximal (45$^\circ$). The solar case is dominated by
$\theta_{12}$, that is large but not maximal.  From a global 3-flavor
analysis of all data one finds the 3$\sigma$ ranges for the mass
differences $\dmsq_{ij} \equiv m^2_i - m^2_j$ and mixing angles
summarized in Table~\ref{tab:mixingparameters}.

\begin{table}[ht]
\caption{\label{tab:mixingparameters}Neutrino mixing parameters from a
global analysis of all experiments 
(3$\sigma$~ranges)~\cite{Gonzalez-Garcia:2002dz}.}
\begin{indented}
\item[]\begin{tabular}{@{}lll}
\br
Observation&Mixing angle&$\Delta m^2$ [meV$^2$]\\
\mr 
Sun, KamLAND & $\theta_{12}=$ 27$^\circ$--42$^\circ$&
$\dmsq_{21} =$ 55--190\\
Atmosphere, K2K  & $\theta_{23} =$ 32$^\circ$--60$^\circ$&
$|\dmsq_{32}|=$ 1400--6000\\
CHOOZ & $\theta_{13} {}<14^\circ$ & 
$\dmsq_{31} \approx \Delta m_{32}^2$\\
\br
\end{tabular}
\end{indented}
\end{table}

A SN core is essentially a neutrino blackbody source, but small
flavor-dependent differences of the fluxes and spectra remain.  We
denote the fluxes of $\nuebar$ and $\nux$ at Earth that would be
observable in the absence of oscillations by $F_\ebar^0$ and $F_x^0$,
respectively.  In the presence of oscillations a $\bar\nu_e$ detector
actually observes
\begin{equation}
F_{\bar{e}}^D(E)  =  \bar{p}^D(E) F_{\bar{e}}^0(E) + 
\left[ 1-\bar{p}^D(E) \right] F_x^0\,,
\label{feDbar}
\end{equation}
where $\bar{p}^D(E)$ is the $\bar\nu_e$ survival probability after
propagation through the SN mantle and perhaps part of the Earth before
reaching the detector.

A significant modification of the survival probability due to the
propagation through the Earth appears only for those combinations of
neutrino mixing parameters shown in Table~\ref{tab:EarthCases}.  The
Earth matter effect depends strongly on two parameters, the sign of
$\dmsq_{32}$ and the value of
$|\theta_{13}|$~\cite{Dighe:1999bi,Dighe:2001rs}.  The ``normal
hierarchy'' corresponds to $m_1<m_2<m_3$, i.e.\ $\dmsq_{32}>0$,
whereas the ``inverted hierarchy'' corresponds to $m_3<m_1<m_2$, i.e.\
$\dmsq_{32}<0$.  Note that the presence or absence of the Earth effect
discriminates between values of $\sin^2 \theta_{13}$ less or greater
than $10^{-3}$, i.e.\ $\theta_{13}$ less or larger than about
$1.8^\circ$.  Thus, the Earth effect is sensitive to values of
$\theta_{13}$ that are much smaller than the current limit.

\begin{table}[ht]
\caption{\label{tab:EarthCases}The Earth effect appears for
the indicated flavors in a SN signal.}
\begin{indented}
\item[]\begin{tabular}{@{}lll}
\br
13-Mixing&Normal Hierarchy&Inverted Hierarchy\\
\mr
$\sin^2\theta_{13} \lsim 10^{-3}$&$\nu_e$ and $\bar\nu_e$
&$\nu_e$ and $\bar\nu_e$\\
$\sin^2\theta_{13} \gsim 10^{-3}$&$\bar\nu_e$&$\nu_e$ \\
\br
\end{tabular}
\end{indented}
\end{table}

Let us consider those scenarios where the mass hierarchy and the value
of $\theta_{13}$ are such that the Earth effect appears for
$\bar\nu_e$.  In such cases the $\bar\nu_e$ survival probability
$\bar{p}^D(E)$ is given by
\begin{equation}
\bar{p}^D \approx \cos^2 \theta_{12} - \sin 2\bar{\theta}_{e2}^\oplus
~ \sin (2\bar{\theta}_{e2}^\oplus - 2\theta_{12}) 
~\sin^2 \left( 12.5\,\frac{\overline{\dmsq_\oplus} L }{E} \right)\,,
\label{pbar}
\end{equation}
where the energy dependence of all quantities will always be implicit.
Here $\bar{\theta}_{e2}^\oplus$ is the mixing angle between $\nuebar$
and $\bar{\nu}_2$ in Earth matter while $\overline{\dmsq_\oplus}$ is
the mass squared difference between the two anti-neutrino mass
eigenstates $\bar{\nu}_1$ and $\bar{\nu}_2$ in units of 
$10^{-5}$eV$^2$, $L$~is
the distance traveled through the Earth in units of 1000~km, and $E$
is the neutrino energy in MeV.  We have assumed a constant matter
density inside the Earth, which is a good approximation for $L<10.5$,
i.e.\ as long as the neutrinos do not pass through the core of the
Earth.

\subsection{Magnitude of Observable Effect at IceCube}

In order to calculate the extent of the Earth effect for IceCube, we
will assume that the relevant mixing parameters are $\Delta
m^2_{12}=6\times10^{-5}~\rm eV^2$ and $\sin^2(2\theta_{12})=0.9$.  We
further assume that the source spectra are given by the functional
form Eq.~(\ref{eq:spectralform}).  The values of the parameters
$\alpha$ and $\langle E \rangle$ for both the $\nuebar$ and
$\bar\nu_x$ spectra are in general time dependent.

\begin{figure}[ht]
\begin{indented}
\item[]
\epsfig{file=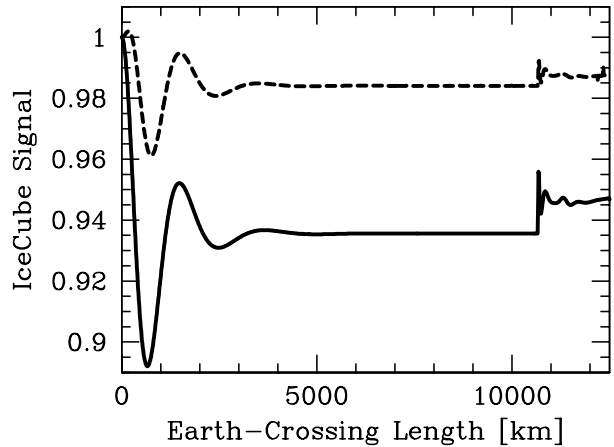,width=8cm}
\end{indented}
\caption{\label{fig:exampleEartheffect}Variation of the expected
IceCube signal with neutrino Earth crossing length $L$ for the assumed
flux and mixing parameters of Table~\ref{tab:ExampleCases}.  The
signal is normalized to 1 when no Earth effect is present, i.e.\ for
$L=0$.  The dashed line is for the case representing the accretion
phase, the solid line for the cooling phase.}
\end{figure}

In Fig.~\ref{fig:exampleEartheffect} we show the variation of the
expected IceCube signal with Earth-crossing length $L$ for the two
sets of parameters detailed in Table~\ref{tab:ExampleCases}.  The
first could be representative of the accretion phase, the second of
the cooling signal.  We use the two-density approximation for the
Earth density profile, where the core has a density of $11.5~\rm
g~cm^{-3}$ and a radius of 3500~km, while the density of the Earth
mantle was taken to be $4.5~\rm g~cm^{-3}$.  We observe that for short
distances, corresponding to near-horizontal neutrino trajectories, the
signal varies strongly with $L$. Between about 3,000 and 10,500~km it
reaches an asymptotic value that we call the ``asymptotic mantle
value.''  For Case~(a), this value corresponds to about 1.5\%
depletion of the signal, whereas for~(b) it corresponds to about 6.5\%
depletion.

\begin{table}[ht]
\caption{\label{tab:ExampleCases}Flux parameters for two
representative cases.}
\begin{indented}
\item[]\begin{tabular}{@{}llllllll}
\br
Example&Phase&$\langle E_{\bar\nu_e}\rangle$&
$\langle E_{\bar\nu_x}\rangle$&
$\alpha_{\bar\nu_e}$&$\alpha_{\bar\nu_x}$&
$\Phi^0_{\bar\nu_e}/\Phi^0_{\bar\nu_x}$&
Asymptotic\\
&&[MeV]&[MeV]&&&&Earth Effect\\
\mr
(a)&Accretion&15&17&4&3&1.5&$-1.5$\%\\
(b)&Cooling  &15&18&3&3&0.8&$-6.5$\%\\
\br
\end{tabular}
\end{indented}
\end{table}

Beyond an Earth-crossing length of $\sim$10,500~km, the neutrinos have
to cross the Earth core with another large jump in density.  The core
effects change the asymptotic mantle value by $\sim 1$\% as can be
seen in Fig.~\ref{fig:exampleEartheffect}. We neglect the core effects
in the following analysis, and the ``asymptotic value'' always refers
to the asymptotic mantle value.

\begin{figure}[ht]
\begin{center}
\begin{indented}
\item[]
\epsfig{file=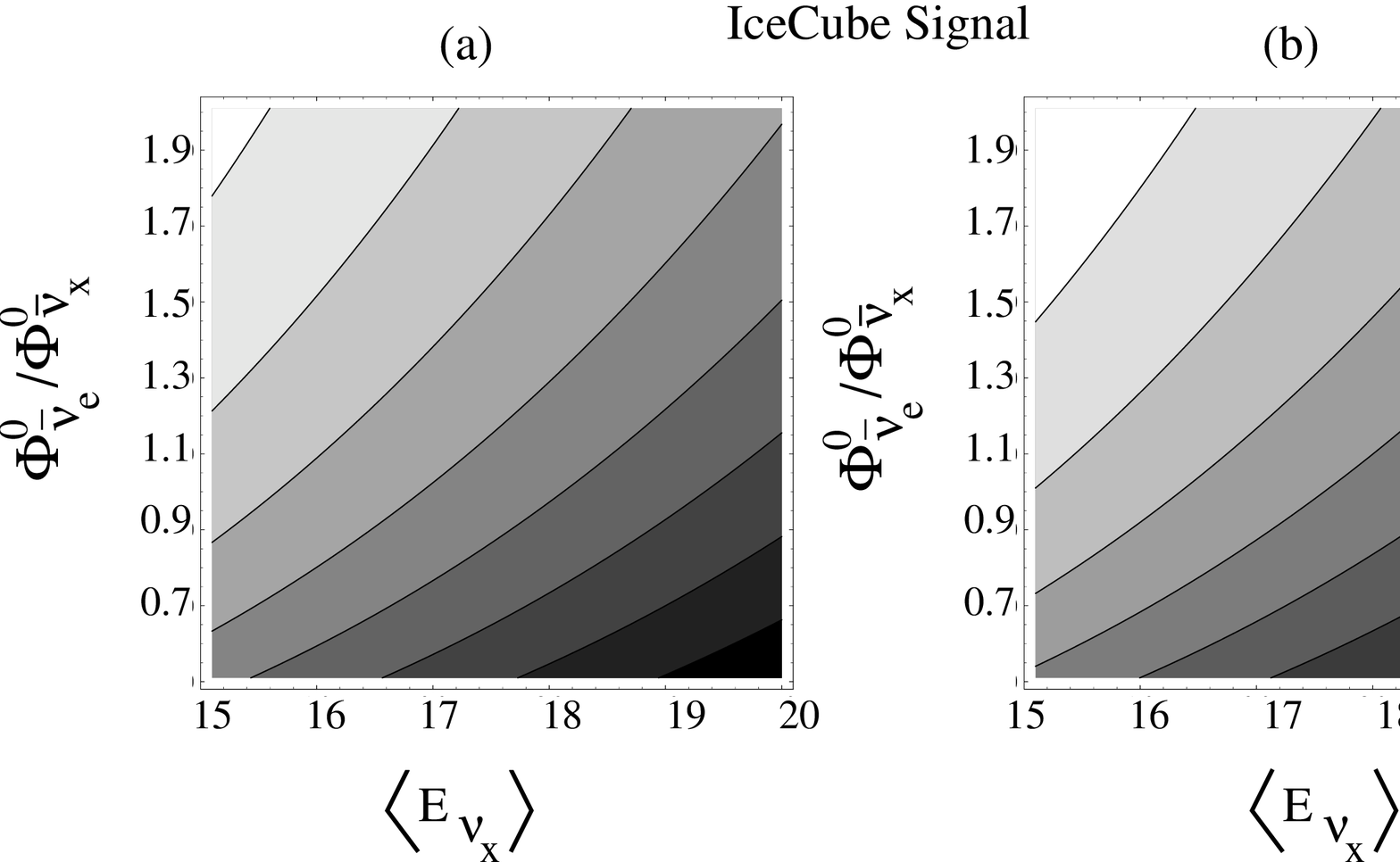,width=13cm}
\end{indented}
\caption{\label{fig:cont} Asymptotic IceCube signal modification by
the Earth effect.  The fixed flux parameters are (a)~$\langle
E_{\bar\nu_e}\rangle=15~\rm MeV$, $\alpha_{\bar\nu_e}=4.0$, and
$\alpha_{\bar\nu_x}=3.0$ and (b)~$\langle E_{\bar\nu_e}\rangle=15~\rm
MeV$, $\alpha_{\bar\nu_e}=\alpha_{\bar\nu_x}=3.0$.  The contours are
equally spaced starting from 1.02 (light) in 0.02 decrements to
smaller values (darker).}
\end{center}
\end{figure}

For the largest part of the sky the Earth effect either appears with
this asymptotic value (``neutrinos coming from below''), or it does not
appear at all (``neutrinos from above''). Therefore, we now focus on
the asymptotic value and study how the signal modification depends on
the assumed flux parameters. In Table~\ref{tab:asymptotic1} we show
the signal modification for $\langle E_{\bar\nu_e}\rangle=15~\rm MeV$,
$\alpha_{\bar\nu_e}=4.0$, and $\alpha_{\bar\nu_x}=3.0$ as a function
of $\langle E_{\bar\nu_x}\rangle$ and the flux ratio
$\Phi^0_{\bar\nu_e}/\Phi^0_{\bar\nu_x}$. In
Table~\ref{tab:asymptotic2} we show the same with
$\alpha_{\bar\nu_e}=\alpha_{\bar\nu_x}=3.0$.  The results are shown in
the form of contour plots in Fig.~\ref{fig:cont}.

\begin{table}[ht]
\vskip24pt
\caption{\label{tab:asymptotic1}Asymptotic IceCube signal modification
by the Earth effect.  The fixed flux parameters are $\langle
E_{\bar\nu_e}\rangle=15~\rm MeV$, $\alpha_{\bar\nu_e}=4.0$, and
$\alpha_{\bar\nu_x}=3.0$.}
\begin{indented}
\item[]\begin{tabular}{@{}lllllll}
\br
Flux ratio&\multicolumn{6}{c}{$\langle E_{\bar\nu_x}\rangle$ [MeV]}\\
$\Phi^0_{\bar\nu_e}/\Phi^0_{\bar\nu_x}$&
15&16&17&18&19&20\\
\mr
2.0 & 1.026 & 1.014 & 1.002 & 0.988 & 0.974 & 0.960 \\
1.9 & 1.023 & 1.011 & 0.999 & 0.985 & 0.971 & 0.956 \\
1.8 & 1.021 & 1.009 & 0.995 & 0.982 & 0.967 & 0.952 \\
1.7 & 1.018 & 1.005 & 0.992 & 0.978 & 0.963 & 0.948 \\
1.6 & 1.015 & 1.002 & 0.988 & 0.974 & 0.959 & 0.944 \\
1.5 & 1.012 & 0.998 & 0.984 & 0.969 & 0.954 & 0.939 \\
1.4 & 1.008 & 0.994 & 0.980 & 0.965 & 0.949 & 0.934 \\
1.3 & 1.004 & 0.990 & 0.975 & 0.960 & 0.944 & 0.928 \\
1.2 & 1.000 & 0.985 & 0.970 & 0.954 & 0.938 & 0.922 \\
1.1 & 0.995 & 0.980 & 0.964 & 0.948 & 0.932 & 0.915 \\
1.0 & 0.989 & 0.974 & 0.957 & 0.941 & 0.925 & 0.908 \\
0.9 & 0.983 & 0.967 & 0.950 & 0.934 & 0.917 & 0.901 \\
0.8 & 0.976 & 0.959 & 0.942 & 0.925 & 0.909 & 0.892 \\
0.7 & 0.967 & 0.950 & 0.933 & 0.916 & 0.899 & 0.883 \\
0.6 & 0.958 & 0.940 & 0.923 & 0.906 & 0.889 & 0.873 \\
0.5 & 0.946 & 0.928 & 0.911 & 0.894 & 0.877 & 0.862 \\
\br
\end{tabular}
\end{indented}
\vskip24pt
\caption{\label{tab:asymptotic2}
Same as Table~\ref{tab:asymptotic1} with
$\alpha_{\bar\nu_e}=\alpha_{\bar\nu_x}=3.0$.}
\begin{indented}
\item[]\begin{tabular}{@{}lllllll}
\br
Flux ratio&\multicolumn{6}{c}{$\langle E_{\bar\nu_x}\rangle$ [MeV]}\\
$\Phi^0_{\bar\nu_e}/\Phi^0_{\bar\nu_x}$&
15&16&17&18&19&20\\
\mr
2.0 & 1.036 & 1.024 & 1.012 & 1.000 & 0.986 & 0.972 \\
1.9 & 1.033 & 1.022 & 1.010 & 0.996 & 0.983 & 0.968 \\
1.8 & 1.031 & 1.019 & 1.006 & 0.993 & 0.979 & 0.964 \\
1.7 & 1.028 & 1.016 & 1.003 & 0.989 & 0.975 & 0.960 \\
1.6 & 1.025 & 1.013 & 0.999 & 0.985 & 0.971 & 0.955 \\
1.5 & 1.022 & 1.009 & 0.995 & 0.981 & 0.966 & 0.951 \\
1.4 & 1.019 & 1.005 & 0.991 & 0.976 & 0.961 & 0.945 \\
1.3 & 1.015 & 1.001 & 0.986 & 0.971 & 0.955 & 0.940 \\
1.2 & 1.010 & 0.996 & 0.981 & 0.965 & 0.949 & 0.933 \\
1.1 & 1.006 & 0.991 & 0.975 & 0.959 & 0.943 & 0.927 \\
1.0 & 1.000 & 0.985 & 0.969 & 0.952 & 0.936 & 0.919 \\
0.9 & 0.994 & 0.978 & 0.961 & 0.945 & 0.928 & 0.911 \\
0.8 & 0.986 & 0.970 & 0.953 & 0.936 & 0.919 & 0.903 \\
0.7 & 0.978 & 0.961 & 0.944 & 0.926 & 0.910 & 0.893 \\
0.6 & 0.968 & 0.950 & 0.933 & 0.916 & 0.899 & 0.882 \\
0.5 & 0.956 & 0.938 & 0.920 & 0.903 & 0.886 & 0.870 \\
\br
\end{tabular}
\end{indented}
\vskip24pt
\end{table}

Even for mildly different fluxes or spectra the signal modification is
several percent, by far exceeding the statistical uncertainty of the
IceCube signal, although the {\em absolute\/} calibration of IceCube
may remain uncertain to within several percent. However, the signal
modification will vary with time during the SN burst. During the early
accretion phase that is expected to last for a few 100~ms and
corresponds to a significant fraction of the overall signal, the
$\bar\nu_x$ flux may be almost a factor of 2 smaller than the
$\bar\nu_e$ flux, but it will be slightly hotter and less
pinched~\cite{Noon03}. This corresponds to Case~(a) above; it is
evident from Fig.~\ref{fig:exampleEartheffect} and
Table~\ref{tab:asymptotic1} that this implies that the Earth effect is
very small. During the Kelvin-Helmholtz cooling phase the flux ratio
is reversed with more $\bar\nu_x$ being emitted than $\bar\nu_e$, but
still with the same hierarchy of energies. This corresponds to
Case~(b); in this case the Earth effect could be about 6\%. This time
dependence may allow one to detect the Earth effect without a precise
absolute detector calibration.

\begin{figure}[ht]
\begin{center}
\begin{indented}
\item[]
\epsfig{file=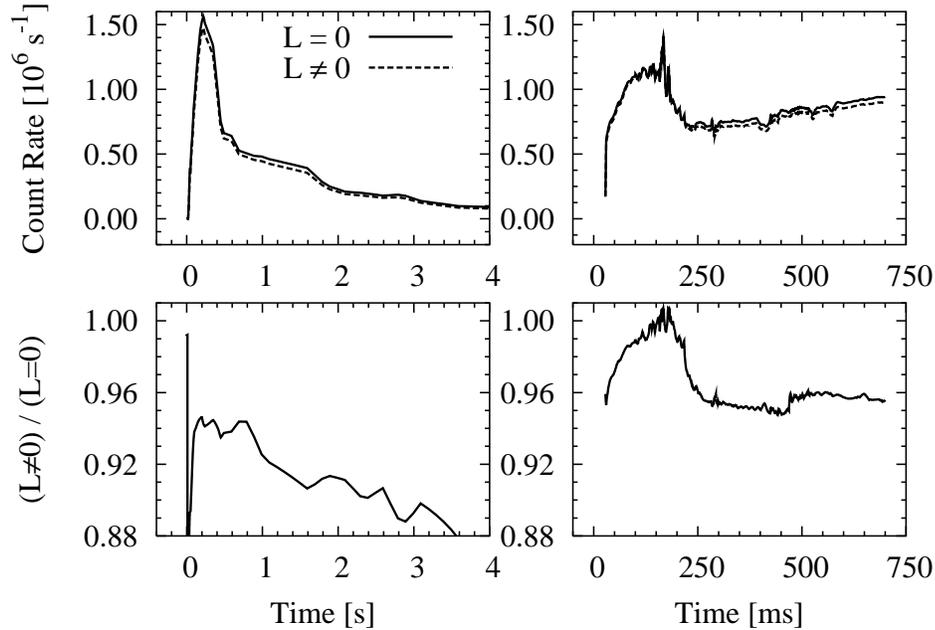,width=13cm}
\end{indented}
\caption{\label{fig:percenteffect} Earth effect in IceCube. The upper
panels show the expected counting rate based on the Livermore (left)
and Garching (right) models, including flavor oscillations. The solid
line is without Earth effect ($L=0$), the dashed line with asymptotic
Earth effect ($L\not=0$). The lower panels show the ratio between
these curves, i.e.\ the ratio of counting rates with/without Earth
effect.}
\end{center}
\end{figure}

In order to illustrate the time dependence of the Earth effect we show
in Fig.~\ref{fig:percenteffect} the expected counting rate in IceCube
for both the Livermore (left panels) and Garching (right panels)
simulations. In the upper panels we show the expected counting rate
with flavor oscillations in the SN mantle, but no Earth effect (solid
lines), or with the asymptotic Earth effect (dashed lines) that
obtains for a large Earth-crossing path. Naturally the differences are
very small so that we show in the lower panels the ratio of these
curves, i.e.\ the expected counting rate with/without Earth effect as
a function of time for both Livermore and Garching.  While for the
Livermore simulation there is a large Earth effect even at early
times, the change from early to late times in both cases is around
4--5\%. Therefore, the most model-independent signature is a time
variation of the Earth effect during the SN neutrino signal.

In order to demonstrate the statistical significance of these effects
we integrate the expected signal for both simulations separately for
the accretion phase and the subsequent cooling phase; the results are
shown in Table~\ref{tab:IntegratedRates}. For both simulations the
Earth effect itself and its change with time is statistically highly
significant. Based on the Livermore simulation, the Earth effect is
much more pronounced than in Garching, the latter involving more
up-to-date input physics for neutrino transport. However, the {\it
difference\/} between the Earth effect during accretion and cooling is
not vastly different between the two simulations.  Recalling that the
absolute detector calibration may be very uncertain so that one has to
rely on the temporal variation of the Earth effect, the difference
between Livermore and Garching becomes much smaller.  We expect that
it is quite generic that the temporal change of the Earth effect is a
few percent of the overall counting rate.

\begin{table}[ht]
\caption{\label{tab:IntegratedRates}
IceCube Cherenkov counts for the numerical SN models.}
\begin{indented}
\item[]\begin{tabular}{@{}lrrrr}
\br
&\multicolumn{2}{c}{Livermore}&\multicolumn{2}{c}{Garching}\\
&\multicolumn{1}{c}{Accretion}&\multicolumn{1}{c}{Cooling}
&\multicolumn{1}{c}{Accretion}&\multicolumn{1}{c}{Cooling}\\
\mr
Integration time [s]&0--0.500&0.500--3&0--0.250&0.250--0.700\\
SN Signal [Counts]\\
\quad No Earth Effect        &519,080&818,043&173,085&407,715\\
\quad Asymptotic Earth Effect&488,093&751,137&171,310&390,252\\
\quad Difference             & 30,987&  66,906&  1,775& 17,463\\
\quad Fractional Difference&$-5.97$\%&$-8.18$\%&$-1.03$\%&$-4.28$\%\\
Background [Counts]&720,000&4,320,000&360,000&648,000\\
$\sqrt{\rm Background}$/Signal&0.16\%&0.25\%&0.35\%&0.20\%\\
\br
\end{tabular}
\end{indented}
\end{table}

\subsection{Super- or Hyper-Kamiokande and IceCube}

One can measure the Earth effect in IceCube only in conjunction with
another high-statistics detector.  We do not attempt to simulate in
detail the SN signal in this other detector but simply assume that it
can be measured with a precision at least as good as in IceCube.  One
candidate is Super-Kamiokande, a water Cherenkov detector that would
measure around $10^4$ events from a galactic SN at a distance of
10~kpc. Therefore, the statistical precision for the total neutrino
energy deposition in the water is around 1\% and thus worse than in
IceCube. Even though Super-Kamiokande will measure a larger number of
Cherenkov photons than IceCube, a single neutrino event will cause an
entire Cherenkov ring to be measured, i.e.\ the photons are highly
correlated.  Therefore, in the estimated statistical $\sqrt{N}$
fluctuation of the signal, the fluctuating number $N$ is that of the
detected neutrinos.  If the future Hyper-Kamiokande is built, its
fiducial volume would be about 30 times that of Super-Kamiokande. In
this case the statistical signal precision exceeds that of IceCube for
the equivalent observable.

We denote the equivalent IceCube signal measured by Super- or
Hyper-Kamiokande as $N_{\rm SK}$ and the IceCube signal as $N_{\rm
IC}$.  If the distances traveled by the neutrinos before reaching
these two detectors are different, the Earth effect on the neutrino
spectra may be different, which will reflect in the ratio $N_{\rm
SK}/N_{\rm IC}$.  Of course, in the absence of the Earth effect this
ratio equals unity by definition.

The geographical position of IceCube with respect to Super- or
Hyper-Kamiokande at a latitude of $36.4^\circ$ is well-suited for the
detection of the Earth effect through a combination of the signals.
Using Fig.~\ref{fig:exampleEartheffect} we can already draw some
qualitative conclusions about the ratio $N_{\rm SK}/N_{\rm IC}$.
Clearly, $N_{\rm SK}/N_{\rm IC}=1$ if neutrinos do not travel through
the Earth before reaching either detector.  If the distance traveled
by neutrinos through the Earth is more that 3000~km for both
detectors, the Earth effects on both $N_{\rm SK}$ and $N_{\rm IC}$ are
nearly equal and their ratio stays around unity.  If the neutrinos
come ``from above'' for SK and ``from below'' for IceCube, or vice
versa, the Earth matter effect will shift this ratio from unity.

\begin{figure}[ht]
\begin{center}
\begin{indented}
\item[]
\epsfig{file=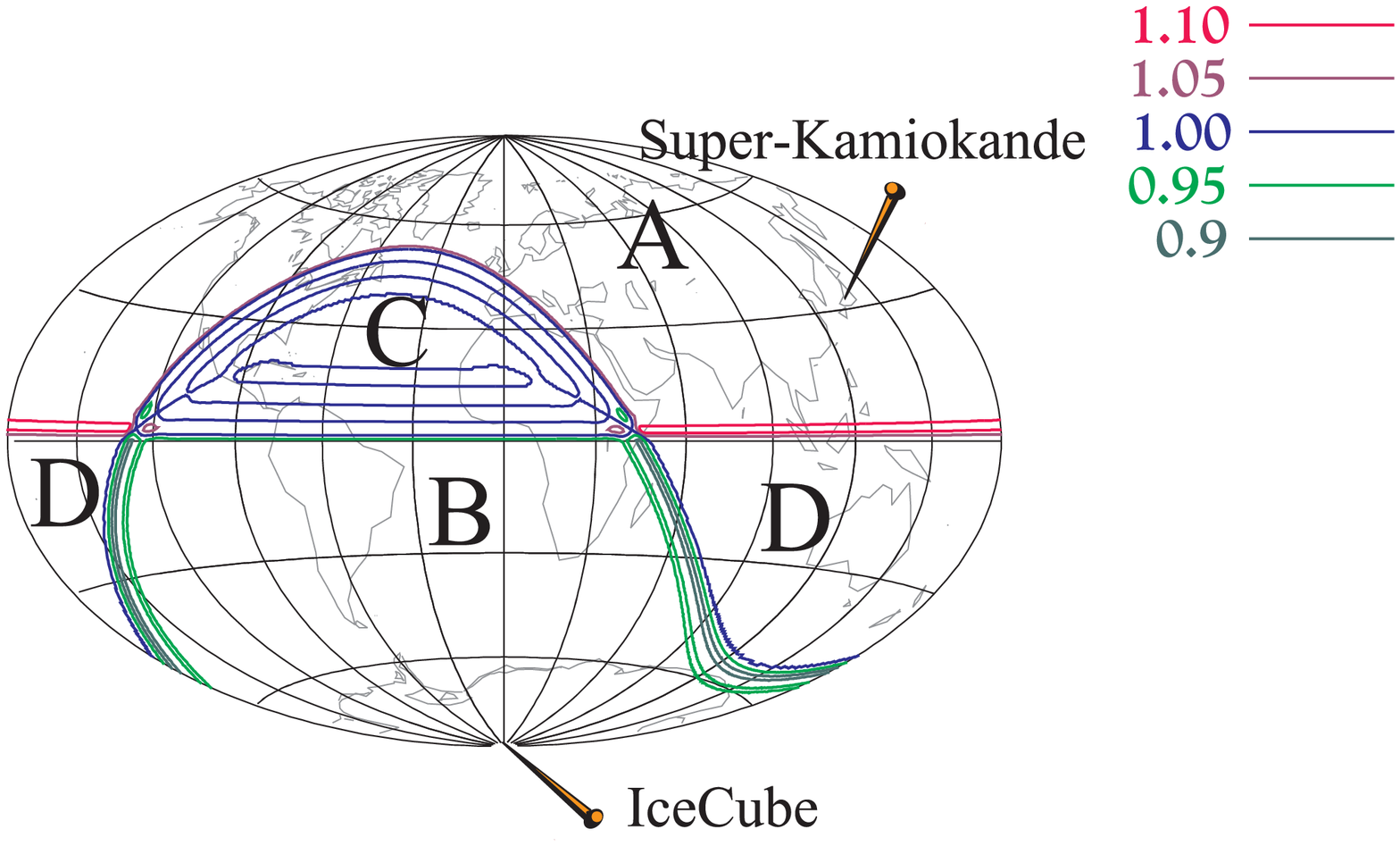,width=13cm}
\end{indented}
\caption{\label{fig:Earthmap}Contours of $N_{\rm SK}/N_{\rm IC}$ on
the map of the sky projected on the Earth. The regions A, B, C, D are
described in Table~\ref{tab:SkyMapRegions}.}
\end{center}
\end{figure}

In Fig.~\ref{fig:Earthmap}, we show contours of $N_{\rm SK}/N_{\rm
IC}$ for the SN position in terms of the location on Earth where the
SN is at the zenith. The map is an area preserving Hammer-Aitoff
projection so that the sizes of different regions in the figure gives
a realistic idea of the ``good'' and ``bad'' regions of the sky.  In
order to generate the contours we use the parameters of Case~(b) in
Table~\ref{tab:ExampleCases} so that the asymptotic suppression of the
signal is about 6.5\%. The sky falls into four distinct regions
depending on the direction of the neutrinos relative to either
detector as described in Table~\ref{tab:SkyMapRegions}.  When the
neutrinos come from above for both detectors (Region~D) there is no
Earth effects. If they come from below in both (Region~C), the Earth
effect is large in both. Depending on the exact distance traveled
through the Earth, the event ratio can be large, but generally
fluctuates around~1.  In the other regions where the neutrinos come
from above for one detector and from below for the other (Regions A
and~B) the relative effect is large.

\begin{table}[ht]
\caption{\label{tab:SkyMapRegions}Regions in Fig.~\ref{fig:Earthmap}
for the Earth effect in IceCube and Super-Kamiokande.}
\begin{indented}
\item[]\begin{tabular}{@{}lllll}
\br
Region&Sky fraction&
\multicolumn{2}{l}{Neutrinos come from}&$N_{\rm SK}/N_{\rm IC}$\\
&&IceCube&Super-K&\\
\mr
A&0.35&below&above&1.070\\
B&0.35&above&below&0.935\\
C&0.15&below&below&Fluctuations around 1 \\
D&0.15&above&above&1\\
\br
\end{tabular}
\end{indented}
\end{table}

\section{Conclusions}
\label{concl}

For assumptions about the flavor-dependent SN neutrino fluxes and
spectra that agree with state-of-the-art studies, the Earth matter
effect on neutrino oscillations shows up in the IceCube signal of a
future galactic SN on the level of a few percent.  If the IceCube
signal can be compared with another high-statistics signal, notably in
Super-Kamiokande or Hyper-Kamiokande, the Earth effect becomes clearly
visible as a difference between the detectors. As one is looking for a
signal modification in the range of a few percent, the absolute
detector calibration may not be good enough in one or both of the
instruments.  However, for typical numerical SN simulations the effect
is time dependent and most notably differs between the early accretion
phase and the subsequent neutron star cooling phase. Therefore, one
would have to search for a temporal variation of the relative detector
signals of a few percent. The large number of optical modules in
IceCube renders this task statistically possible.  In fact depending
on the differences in flavor-dependent fluxes, the statistical
accuracy of Super-Kamiokande may turn out to be the limiting
factor. This limitation is not significant for Hyper-Kamiokande.

The unique location of IceCube in Antarctica implies that for a large
portion of the sky this detector sees the SN through the Earth when
Super- and Hyper-Kamiokande sees it from above, or the other way
round, i.e.\ the chances of a relative signal difference between the
detectors are large. If both detectors were to see the SN from above
there would be no Earth effect to detect.

Assuming that the magnitude of the mixing angle $\theta_{13}$ can be
established to be large in the sense of
$\sin^2\theta_{13}\gsim10^{-3}$ by a long-baseline
experiment~\cite{Barger:2000nf,Cervera:2000kp,Freund:2001ui},
observing the Earth effect in SN anti-neutrinos implies the normal
mass hierarchy.  On the other hand, if $\sin^2\theta_{13}\lsim10^{-3}$
has been established, the Earth effect is unavoidable. Not observing
it would imply that the primary SN neutrino fluxes and spectra are
more similar than indicated by state-of-the-art numerical simulations.

If $\sin^2\theta_{13}\gsim10^{-3}$ is known, and we do not observe the
Earth effect, it still does not prove the inverted mass hierarchy.  It
could also mean that we do not properly understand the
flavor-dependent source fluxes and spectra. Therefore, even if
$\sin^2\theta_{13}\gsim10^{-3}$ is known, our method only allows one
to detect the normal mass hierarchy, it does not strictly allow one to
exclude it. As far as neutrino parameters are concerned, only a
positive detection of the Earth effect would count for much.  Of
course, a normal mass hierarchy and $\sin^2\theta_{13}\gsim10^{-3}$ is
certainly a plausible scenario so that expecting a positive
identification of the Earth effect is not a far-fetched possibility.

In summary, even though galactic SNe are rare, the anticipated
longevity of IceCube and the long-term neutrino program at Super- or
Hyper-Kamiokande imply that detecting the Earth effect in a SN
neutrino burst is certainly a distinct possibility. This could
identify the normal neutrino mass hierarchy, a daunting task at
long-baseline
experiments~\cite{Barger:2000nf,Cervera:2000kp,Freund:2001ui}.  Given
the difficulty of pinning down the mass hierarchy at long-baseline
experiments, both IceCube and Super- or Hyper-Kamiokande should take
all instrumental and experimental steps required to ensure the
feasibility of a high-statistics simultaneous SN observation.

\section*{Acknowledgments}

This work was supported, in part, by the Deutsche
Forschungsgemeinschaft under grant No.\ SFB-375 and by the European
Science Foundation (ESF) under the Network Grant No.~86 Neutrino
Astrophysics. We thank Francis Halzen and Robert Buras for helpful
comments on an early version of this manuscript.

\newpage

\section*{References}

\end{document}